\documentclass[superscriptaddress,secnumarabic,twocolumn,
 amssymb,amsmath,nobibnotes,aps,prd,showkeys,showpacs]{revtex4}
\usepackage{graphicx,epsfig,amssymb,amsmath,amstext}

\usepackage{graphicx}
\usepackage{bm}

\begin{document}

\title{The non-integrability of the Zipoy-Voorhees metric}

\author{Georgios Lukes-Gerakopoulos}
\affiliation{Theoretical Physics Institute, University of Jena, 07743 Jena,
Germany}
\email{gglukes@gmail.com}

\begin{abstract}
 
The low frequency gravitational wave detectors like eLISA/NGO will give us the
opportunity to test whether the supermassive compact objects lying at the
centers of galaxies are indeed Kerr black holes. A way to do such a test is to
compare the gravitational wave signals with templates of perturbed black hole
spacetimes, the so-called bumpy black hole spacetimes. The Zipoy-Voorhees (ZV)
spacetime (known also as the $\gamma$ spacetime) can be included in the bumpy
black hole family, because it can be considered as a perturbation of the
Schwarzschild spacetime background. Several authors have suggested that the ZV 
metric corresponds to an integrable system. Contrary to this integrability
conjecture, in the present article it is shown by numerical examples that in
general ZV belongs to the family of non-integrable systems.
\end{abstract}

\maketitle

\section{Introduction}\label{sec:Intro}

It is possible that in the next decade the European New Gravitational Wave
Observatory (NGO) \cite{Amaro12a,Amaro12b} will be launched. The NGO will
allow the tracing of the spacetime around supermassive compact objects
($10^5-10^7~M_{\odot}$), lying at the centers of galaxies, by detecting the
gravitational waves emitted by an inspiraling much less massive compact object,
e.g. a stellar mass black hole or a neutron star. The motion in such a binary
system is known as Extreme Mass Ratio Inspiral (EMRI). We anticipate that these
supermassive compact objects are black holes and the spacetime around them is
described by the Kerr metric.

However, we must check this hypothesis. An experimental test is to extract
information from gravitational waves emitted by an EMRI system. Ryan showed
\cite{Ryan} that we can extract the multipole moments of the central body from
the gravitational wave signal (see also \cite{Li08,Pappas12a}). Thus, any
non-Kerr multipole moments should be encoded in the waves. This was basically
the idea Collins and Hughes originally conceived and stated in \cite{Collins04}.
They, therefore, constructed a perturbed black hole which remained stationary
and axisymmetric, and named it a ``bumpy'' black hole. According to their
original idea, we should be able to measure deviations of the multipole moments
of this perturbed black hole from those of a Kerr black hole in cases of EMRIs.
But instead of following this approach, they preferred to explore the bumpiness
of the perturbed spacetimes by measuring the periapsis precession
\cite{Collins04}. Their original idea was implemented later by Vigeland
\cite{Vigeland10a}.

Since Collins and Hughes' work, various approaches \cite{Glamped06,
Barausse07,Gair08,Apostolatos09,Vigeland10b,Lukes10,Vigeland11,Gair11},
have been applied in order to identify the imprints of a perturbed spacetime
background on the gravitational wave signal in the case of an EMRI system. On
the other hand, the gravitational wave spectra are not the only spectra which
can probe the bumpiness of a black hole. Electromagnetic spectra have been
suggested in several works \cite{JohanPsal,Bambi,BamBar11,Pappas12b} as a
plausible alternative. For more information about bumpy black hole detection
methods see \cite{Bambi11,Johannsen12}.     

In previous works, we approached the bumpy black hole topic by studying the
non-linear dynamics of the geodesic motion on the corresponding perturbed
background \cite{Lukes10,Contop11,Contop12}. Phenomena related to non-linear
dynamics appear in the bumpy black hole spacetimes due to the fact that the
corresponding (stationary and axisymmetric) systems are missing a fourth
integral of motion analogous to that of the Kerr spacetime, i.e. the Carter
constant \cite{Carter68}. In fact, the existence of this extra symmetry,
associated with the Carter constant, is the reason that the Kerr system is
integrable.

The Carter constant seems to hold only for certain types of systems (see
\cite{Will09,Markakis12} and references therein) which include the Kerr system.
However, it is not typical to find spacetimes which are integrable.

The Zipoy-Voorhees (ZV) metric \cite{Zipoy66,Voorhees70}, known also as the
$\gamma$ metric \cite{EspositoWitten75}, has been conjectured to correspond to
an integrable system. In particular, in \cite{Sota96} the authors examined the
ZV spacetime numerically and did not find signs of chaos. A new numerical
investigation by Brink in \cite{Brink08} gave the same result, which led Brink
to conjecture that the ZV spacetime is integrable. Hence, she tried to find the
missing integral of motion \cite{Brink08}. The question of the ``missing''
integral was recently addressed also in \cite{KrugMat11}. In this article the
authors initially conjectured that the ZV is integrable in order to search for
the missing integral, however their investigation led them to the opposite
direction. Namely, they proved the nonexistence of certain types of integrals of
motion in a specific case of the ZV metric.

The integrability conjecture is certainly true in two special cases, i.e. when a
``free'' parameter of the metric gives the Minkowski metric or the Schwarzschild
metric. But, in the present article it is shown that this does not hold in
general. In general, the phase space of the ZV system has all the features of a
perturbed system, like chaotic layers, Birkhoff chains etc. These features are
found by using Poincar\'{e} sections and by employing the so-called rotation
number indicator \cite{Laskar93,Contop97,Voglis98}. Thus, by numerical examples
it will be shown that in general the ZV metric corresponds to a non-integrable
system.

The paper is organized as follows. Section \ref{sec:ThEl} summarizes some 
basic elements regarding the ZV metric, the geodesic motion in the ZV spacetime
background, as well as some essential theoretical elements regarding Hamiltonian
nonlinear dynamics. Numerical examples of the ZV non-integrability are presented
in section \ref{sec:NumRes}. Section \ref{sec:ConDis} summarizes and discusses
the main results of the present work. The accuracy of the integration method
used to calculate the geodesic orbits is discussed in appendix \ref{sec:NumAc}.

\section{Theoretical elements}\label{sec:ThEl}

\subsection{The Zipoy-Voorhees spacetime}\label{subsec:ZVs}

The Zipoy-Voorhees metric \cite{Zipoy66,Voorhees70} describes a two parameter
family of Weyl class spacetimes, which are static, axisymmetric and
asymptotically flat vacuum solutions of the Einstein equations. The ZV line
element in prolate spheroidal coordinates is given by
\begin{equation}\label{eq:leZVpsc}
 ds^2=g_{tt}~dt^2+g_{xx}~dx^2+g_{yy}~dy^2+g_{\phi\phi}~d\phi^2,
\end{equation}
where the metric elements are
\begin{eqnarray}\label{eq:meZVpsc}
 g_{tt} &=&-\left(\frac{x-1}{x+1}\right)^\delta, \nonumber \\
 g_{\phi\phi} &=& k^2~\left(\frac{x+1}{x-1}\right)^\delta (x^2-1)(1-y^2),
 \nonumber \\
 \\
 g_{xx} &=& k^2~\left(\frac{x+1}{x-1}\right)^\delta~
 \left(\frac{x^2-1}{x^2-y^2}\right)^{\delta^2}\frac{x^2-y^2}{x^2-1}, \nonumber\\
 g_{yy} &=& k^2~\left(\frac{x+1}{x-1}\right)^\delta
 \left(\frac{x^2-1}{x^2-y^2}\right)^{\delta^2}\frac{x^2-y^2}{1-y^2},
 \nonumber 
\end{eqnarray} 
and $k=M/\delta$ is the ratio of the source mass $M$ and of the ``oblateness''
parameter $\delta$. The ZV quadrapole moment is
$\displaystyle{\cal Q}=M^3\frac{\delta(1-\delta^2)}{3}$ \cite{Herrera99}.
If $\delta=1$ the ZV metric (\ref{eq:meZVpsc}) describes the standard
Schwarzschild spacetime and the corresponding central object is spherically
symmetric $({\cal Q}=0)$. If $\delta>1$ the ZV metric describes a spacetime
around a central object more oblate than a Schwarzschild black hole; if
$0<\delta<1$ the central object is more prolate. Finally, if $\delta=0$ we get
the Minkowski flat spacetime.
 
The ZV metric takes a more familiar form 
\cite{EspositoWitten75}\footnote{In \cite{EspositoWitten75} the symbol
$\gamma$ was used instead of $\delta$} by using the transformation
\begin{equation}\label{eq:spcTrBL}
 x=\frac{r}{k}-1,~~~ y=\cos{\theta}.
\end{equation}
Then, the metric element (\ref{eq:leZVpsc}) is
\begin{eqnarray}\label{eq:leGamma}
 ds^2 &=& -F dt^2+F^{-1}[G~dr^2+H~d\theta^2 \nonumber \\
      &+& (r^2-2k~r)\sin^2\theta d\phi^2],
\end{eqnarray}
where
\begin{eqnarray} \label{eq:meGamma}
 F &=& \left(1-\frac{2k}{r}\right)^\delta, \nonumber\\
 G &=& \left(\frac{r^2-2k~r}{r^2-2k~r+k^2 \sin^2 \theta}\right)^{\delta^2-1}, \\
 H &=& \frac{(r^2-2k~r)^{\delta^2}}{(r^2-2k~r+k^2 \sin^2 \theta)^{\delta^2-1}}.
\nonumber
\end{eqnarray}
From this formulation, it is easy to check that for $\delta=1$ we get the
Schwarzschild spacetime around a central object of mass $k$ \footnote{The
geometric units are used throughout the article, i.e. the speed of light in
vacuum and the gravitational constant are set equal to unity.}. This means that
the parameter $\delta$ also ``measures'' how much more (or less) mass
$M=\delta~k$ the ZV central object has compared with a Schwarzschild black hole
of mass $M=k$. 

For $0<\delta\neq 1$ the event horizon is broken and a curvature singularity
appears along the line segment $\rho=0,~|z|<k$ \cite{Kodama03}. Thus, the ZV
spacetimes in general describe naked singularities (for further information see
\cite{Herrera99,Kodama03,Papadopoulos81}). On the other hand, there are no signs
of closed timelike curves ($g_{\phi\phi} \ge 0,~\forall~x,~y$).

The parameter $\delta$ has one more property of present interest: from a
dynamical point of view $\delta$ can be seen as a perturbation parameter of the
Schwarzschild spacetime. When the value of $\delta$ departs from $1$, the
spherical symmetry is broken and the new axisymmetric spacetime needs not
possess a Carter-like constant. In fact, some indication of this absence was
found in \cite{KrugMat11}, where for the $\delta=2$ case the authors proved that
a broad category of integrals of motion has to be excluded from being the
integral that would make ZV integrable. In section \ref{sec:NumRes} it is shown
that indeed such an integral cannot exist, because for $\delta=2$ there are
geodesic orbits which are chaotic. Similar results are found for other
representative values of $\delta$ as well.

\begin{figure*}[htp]
 \centerline{\includegraphics[width=0.8\textwidth] {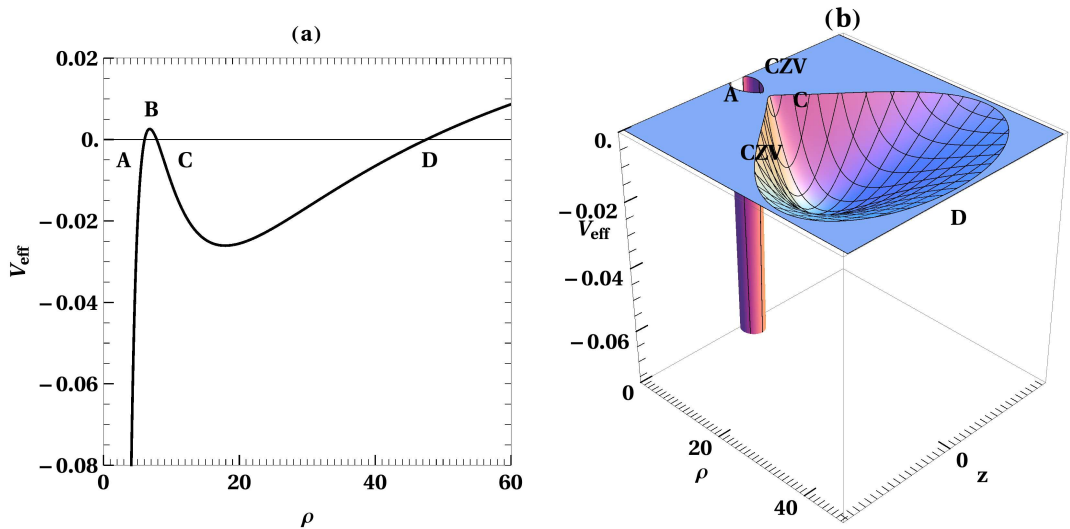}}
 \caption{(a) The black curve shows the effective potential $V_{\textrm{eff}}$
 for $\delta=2,~k=1,~E=0.97,~L_z=7.6,~z=0$ along the $\rho$ axis. The letters A,
 C, D indicate the positions of the $V_{\textrm{eff}}$ roots for $z=0$, while
 the black horizontal line shows the zero value of the $V_{\textrm{eff}}$ axis.
 The letter B indicates the position of a local maximum of the
 $V_{\textrm{eff}}$, which separates the plunging orbits from the bounded
 non-plunging. (b) The effective potential cut along the $V_{\textrm{eff}}=0$
 plane. The A, C, D show the same as in (a) and the CZV corresponds to the
 curves of zero velocity.  
 }
 \label{FigVeff}
\end{figure*}

\subsection{Geodesic motion in the Zipoy Voorhees spacetime background}
\label{subsec:GMZV}

The equations of geodesic motion of a ``test'' particle of rest mass $\mu$ in a
spacetime given by the metric $g_{\mu\nu}$ are produced by the Lagrangian
function
\begin{equation} \label{eq:LagDef}
L=\frac{1}{2}~\mu~g_{\mu\nu}~ \dot{x}^{\mu} \dot{x}^{\nu}
\end{equation}
where the dot denotes derivation with respect to proper time and the Greek
indexes stand for spacetime coordinates. The Lagrangian (\ref{eq:LagDef}) has a
constant value $L=-\frac{\mu}{2}$ along a geodesic orbit, due to the 
$g_{\mu\nu}~\dot{x}^{\mu}\dot{x}^{\nu}=-1$ constraint.

Since the ZV spacetime is axisymmetric and stationary the corresponding momenta
\begin{equation}\label{eq:MomDef}
 p_\nu=\partial{L}/\partial{\dot{x}^\nu}
\end{equation}
are conserved. These are the specific energy \footnote{Per unit mass.}
\begin{equation} \label{eq:EnCon}
E=-\frac{\partial L}{\partial \dot{t}}/\mu=-g_{tt}\dot{t},
\end{equation}
and the specific azimuthal component of the angular momentum
\begin{equation} \label{eq:AnMomCon}
L_z =\frac{\partial L}{\partial \dot{\varphi}}/\mu=g_{\phi\phi}\dot{\phi}~~.
\end{equation}
For brevity, these two integrals are hereafter simply referred to as the energy
$E$ and the angular momentum $L_z$.

The $z$ index refers to the cylindrical $(\rho,z)$ coordinate system used in all
figures. The transformation equations from spheroidal prolate to cylindrical
coordinates is
\begin{equation}\label{eq:spcTcc}
 \rho=k\sqrt{(x^2-1)(1-y^2)},~~~z=k~x~y~~.
\end{equation}
The set of coordinates $(\rho,z)$ describes the meridian plane. Due to the two
integrals of motion (\ref{eq:EnCon}), (\ref{eq:AnMomCon}) we can restrict our
study to this plane. We just have to reexpress (\ref{eq:EnCon}),
(\ref{eq:AnMomCon}) to get $\dot{t}$ and $\dot{\phi}$ as functions of $E$ and
$L_z$, i.e.
\begin{equation} \label{eq:tphi}
 \dot{t}=-E/g_{tt},~~\dot{\phi}=L_z/g_{\phi\phi}~~,
\end{equation}
and then substitute $\dot{t}$, $\dot{\phi}$ into the two remaining equations of
motion. Then, from the original set of $4$ coupled second order ordinary
differential equations (ODEs), we arrive at a set of $2$ coupled ODEs.     

The motion on the meridian plane satisfies yet another constraint. If we use the 
transformation (\ref{eq:spcTcc}) on the metric (\ref{eq:meZVpsc}), substitute
(\ref{eq:tphi}) and replace the Lagrangian function with its constant value, we
get
\begin{equation} \label{eq:NewVeff}
\frac{1}{2} (\dot{\rho}^2 + \dot{z}^{2}) + V_{\textrm{eff}} (\rho, z)=0,
\end{equation}
where
\begin{eqnarray}\label{eq:Veff}
 V_{\textrm{eff}} &=& \frac{1}{2}Z\left(1+\frac{E^2}{g_{tt}}+
 \frac{L_z^2}{g_{\phi\phi}}\right), \\
 Z &=& \left(\frac{r_{+}+r_{-}-2k}{r_{+}+r_{-}+2k}\right)^\delta
   \left(\frac{1}{2}+\frac{1}{4}\left(\frac{r_{+}}{r_{-}}
    +\frac{r_{-}}{r_{+}}\right)\right)^{-\delta^2}, \nonumber \\
 r_{\pm} &=& \sqrt{\rho^2+(z\pm k)^2}. \nonumber
\end{eqnarray}
The $V_{\textrm{eff}}$ corresponds to a Newtonian-like two-dimensional effective
potential (Fig. \ref{FigVeff}). The roots ($V_{\textrm{eff}}=0$) of this
effective potential (Fig. \ref{FigVeff}a) produce a curve on the meridian plane
$(\rho,~z)$ called the curve of zero velocity (CZV) (Fig. \ref{FigVeff}b). This
nomenclature follows from noting that whenever $V_{\textrm{eff}}=0$ the
constraint (\ref{eq:NewVeff}) gives $\dot{\rho}=\dot{z}=0$, which means that
whenever a geodesic orbit reaches the CZV the velocity component in the
$\rho,~z$ plane becomes equal to zero.

On the meridian plane the CZV determines the set of initial conditions of
geodesic orbits not escaping to infinity. These bounded orbits can either plunge
to the central ZV compact object or revolve around it. In Fig. \ref{FigVeff}a
the plunging orbits lie before point A, while the non-plunging orbits lie
between the points C and D. In Fig. \ref{FigVeff}b the point A lies on the CZV
containing all the plunging orbits, while C and D lie on the CZV containing the
non-plunging orbits. This separation results from the fact that the local
maximum of the effective potential at point B is greater than 0. If the angular
momentum is reduced below $L_z=7.58$ for $\delta=2,~E=0.97$ (Fig.
\ref{FigVeff}a), the maximum B drops below $0$ and a saddle point appears. This
saddle point corresponds to an unstable periodic orbit similar to the unstable
periodic orbit denoted with the letter ``x'' in \cite{Contop11,Contop12}, this
notation is adopted in the present article as well. 

In the integrable case $\delta=1$ (Schwarzschild metric) from such an
unstable orbit ``x'' emanate the branches of the separatrix manifold, which
separate the plunging from the non-plunging orbits. When the system is
non-integrable, instead of the separatrix manifold, there are \emph{asymptotic
manifolds} emanating from the ``x'' orbit. ``x'' is formally called Lyapunov
orbit (LO) \cite{Contop90}. Every orbit crossing the border defined by a LO
``escapes'', which in our case means that every orbit which passes the LO while
moving towards the central object, will plunge to the central object. On the
other hand, for certain energy and angular momentum the ``x'' orbit changes its
stability to indifferently stable and becomes the innermost stable circular
orbit (ISCO).

In order to locate the ISCO, it is more convenient  for the algebraic 
manipulation to use the potential $V=2~V_\textrm{eff}/Z$, instead of the
effective potential (\ref{eq:Veff}). For the ISCO radius $x_\textrm{ISCO}$ the
potential $V$ and its two first derivatives with respect to $x$ are equal to
zero, i.e. $V|_{x_\textrm{ISCO}}=\frac{\partial V}{\partial x}|_{x_\textrm{ISCO}}=
\frac{\partial^2 V}{\partial x^2}|_{x_\textrm{ISCO}}=0$. Then we find
\begin{eqnarray} \label{eq:Isco}
 x_\textrm{ISCO} &=& 3 \delta+\sqrt{5 \delta^2-1}, \nonumber \\ 
 E_\textrm{ISCO} &=& \sqrt{-g_{tt}|_{x_\textrm{ISCO}}
 \frac{x_\textrm{ISCO}-\delta}{x_\textrm{ISCO}-2\delta}}, \\
 L_{z_\textrm{ISCO}} &=& \sqrt{\frac{\rho^2}{g_{tt}|_{x_\textrm{ISCO}}}
(1+\frac{E^2}{g_{tt}|_{x_\textrm{ISCO}}})} \nonumber
\end{eqnarray}
for $\delta \geq 1/\sqrt{5}$. For $\delta < 1/\sqrt{5}$, on the other hand the 
ZV spacetime ceases to have an ISCO. From a dynamical point of view, below this
limit we may consider the ZV rather as a perturbation of the Minkowski
($\delta=0$) than of the Schwarschild ($\delta=1$) spacetime. The ISCO in the ZV
spacetime was studied recently in \cite{Chowdhury12}, where the authors
investigated the possible observational differences between the properties of
accretion disks in the case of a ZV spacetime and a black hole spacetime.
       
\subsection{Nonlinear Hamiltonian dynamics}\label{subsec:HamDyn}

By simply applying a Legendre transformation
\begin{equation}\label{eq:LegTr}
 H=p_\mu \dot{x}^\mu-L
\end{equation}
on the Lagrangian function (\ref{eq:LagDef}), we get the Hamiltonian function
\begin{equation}\label{eq:HamDef}
 H=\frac{1}{2~\mu}g^{\mu\nu} p_\mu p_\nu
\end{equation}
where the momenta $p_\mu$ are given by (\ref{eq:MomDef}) and
$p^\nu=\mu \dot{x}^\nu$. It follows immediately that the ZV system is an
autonomous system ($\frac{d H}{d \tau}=\frac{\partial H}{\partial \tau}=0$),
thus the Hamiltonian function is an integral of motion and equal to
$H=-\mu/2$ (because $g^{\mu\nu} p_\mu p_\nu=-\mu^2$) 

As already mentioned in section \ref{subsec:GMZV}, the study of an axisymmetric
and stationary system can be reduced to the meridian plane. Therefore, the
system is reduced to a Hamiltonian system with two degrees of freedom. In the
case of the Schwarzschild metric ($\delta=1$), spherical symmetry introduces an
extra integral of motion, namely the total angular momentum. This integral is
the reason why the Schwarzschild metric corresponds to an integrable system. In
this case bounded non-plunging orbits oscillate in both degrees of freedom with
two characteristic frequencies, denoted hereafter $\omega_1$ and $\omega_2$. The
motion is restricted on a two-dimensional torus, called \emph{invariant torus}.
The type of motion on the torus depends on the ratio $\omega_1/\omega_2$. If the
ratio is a rational number, the motion is periodic. The corresponding torus is
called resonant and it hosts infinitely many periodic orbits with the same
frequency ratio. On the other hand, if $\omega_1/\omega_2$ is  irrational, the
motion is called \emph{quasiperiodic}. In that case, one single orbit densely
covers the whole non-resonant torus.

In general, resonant and non-resonant tori align around a central periodic orbit
forming the so-called \emph{tori foliation}. If we move away from the central
periodic orbit along any radial direction, we encounter tori with varying
characteristic frequencies $\omega_1,~\omega_2$. In fact, the ratio 
$\omega_1/\omega_2$ changes along the radial direction in the same way as the
rational and the irrational numbers interchange along a real axis. 

Now, if an integrable system (like Schwarzschild) is perturbed, the transition
from integrability to non-integrability follows two basic theorems, namely the
Kolmogorov-Arnold-Moser (KAM) theorem \cite{KAM} and the Poincar\'{e}-Birkhoff
theorem \cite{PoinBirk}.

According to the KAM theorem, for small perturbations, most non-resonant
invariant tori of the integrable system survive deformed in the perturbed system.
The new deformed tori are called KAM tori. 

On the other hand, according to the Poincar\'{e}-Birkhoff theorem, from the
infinitely many periodic orbits on a resonant torus of the integrable system 
only an even number survive in the perturbed system. Half of these surviving
periodic orbits are stable and the other half unstable.

\begin{figure}[htp]
 \centerline{\includegraphics[width=0.3\textwidth] {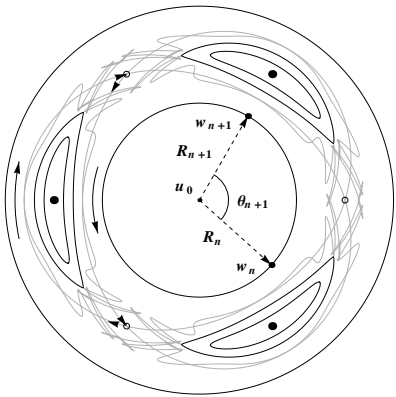}}
 \caption{A schematic representation of a surface of section of a non-integrable
 Hamiltonian system of two degrees of freedom. The figure shows a Birkhoff
 chain at the resonance $1/3$ between two KAM curves (homocentric cyclic curves
 around the central periodic point $\mathbf{u}_0$). The Birkhoff chains consists
 of 3 unstable points (open circles) and 3 stable points (filled circles). An
 island of stability is formed around each of the 3 stable points (elliptic
 curves). From the unstable points emanate asymptotic curves (gray curves). The
 arrowheads on the asymptotic curves indicate the stable and the unstable branch.
 The curved arrows indicate the flow in the phase space. The points
 $\mathbf{w}_n,~\mathbf{w}_{n+1}$ indicate two successive crossings of the
 quasiperiodic orbit forming the KAM curve. $\theta_{n+1}$ is the angle between
 the position vectors $\mathbf{R}_n,~\mathbf{R}_{n+1}$ (with respect to
 $\mathbf{u}_0$) of the points  $\mathbf{w}_n,~\mathbf{w}_{n+1}$ respectively.
 }
 \label{FigSurSecNI}
\end{figure}

The structure of the phase space is revealed conventionally by the use of a
2-dimensional Poincar\'{e} surface of section. On a surface of section, the
invariant tori correspond to closed curves, such curves are shown in Fig.
\ref{FigSurSecNI} (schematic). For weakly perturbed systems the KAM curves on
the surface of section are closed invariant curves around a stable fixed point
in the center (like $\mathbf{u}_0$ in Fig. \ref{FigSurSecNI}). This structure is
hereafter called the \emph{main island of stability}.  

However, the details of the surface of section of a non-integrable Hamiltonian
system close to resonances are quite different from those of integrable systems.
The stable surviving periodic orbits are depicted as stable points
on a surface of section (filled circles in Fig. \ref{FigSurSecNI}) and these
points are surrounded by other regular orbits forming the secondary islands of
stability (Fig. \ref{FigSurSecNI}). Between these islands lie unstable points
(open circles in Fig. \ref{FigSurSecNI}), which correspond to unstable periodic
orbits. From the unstable periodic orbits emanate their asymptotic manifolds,
yielding asymptotic curves on the surface of section (gray curves of Fig.
\ref{FigSurSecNI}). There are two types of asymptotic curves: stable and
unstable. A stable (unstable) asymptotic manifold cannot cross itself or other
stable (unstable) asymptotic manifolds. Initial conditions on a stable
(unstable) manifold tend asymptotically to the periodic orbit in the inverse
(direct) flow of the time parameter. The latter property of asymptotic
manifolds, combined with the non-crossing property produce on the surface of
section oscillations of the  manifolds, causing the so-called homoclinic
``chaos'' effect. Due to this, if we start with initial conditions within the
domain crossed by manifolds, their subsequents give the impression of scattered
points, which are a signature of chaotic motion on a surface of section.

Besides the use of a surface of section, another tool to study non-integrable
systems is the \emph{rotation number} $\nu_\theta$. The rotation number has been
proved to be an efficient indicator of chaos \cite{Laskar93,Contop97,Voglis98}
(see \cite{Contop02} for review). A simple way to evaluate the rotation number
is the following. First we identify the central periodic orbit ($\mathbf{u}_0$
in Fig. \ref{FigSurSecNI}) of the main island of stability. Then, we define the
position vector
\begin{equation} \label{eq:PosVec}
 \mathbf{R}_n=\mathbf{w}_n-\mathbf{u}_0
\end{equation}
of the n-th crossing $\mathbf{w}_n$ of the orbit through a surface of section
with respect to the central periodic orbit $\mathbf{u}_0$ (dashed vectors in Fig.
\ref{FigSurSecNI}). We then find the angle between two successive vectors
$\theta_{n+1}\equiv angle (\mathbf{R}_{n+1},\mathbf{R}_n)$ (called the
rotation angle) and we evaluate the rotation number
\begin{equation} \label{eq:RotNum}
 \nu_\theta=\frac{1}{2 \pi N}\sum_{n=1}^N\theta_n~~.
\end{equation}
In the limit $N\rightarrow \infty$, the rotation number corresponds to the
frequency ratio $\omega_1/\omega_2$.

If we plot the rotation number as a function of the distance of initial
conditions from the central periodic orbit $\mathbf{u}_0$ of the main island of
stability along a particular direction we obtain the so-called \emph{rotation
curve}. For integrable systems, like the Schwarzschild metric, the rotation
curve is a smooth and strictly monotonic function. When the system is perturbed,
the rotation curve ceases to be smooth, and it is only approximately monotonic.
In fact, the imprints of chaos are found in the rotation curve, as demonstrated
by several examples in section \ref{sec:NumRes}.

\section{Numerical results} \label{sec:NumRes}

\begin{figure}[htp]
 \centerline{\includegraphics[width=0.5\textwidth] {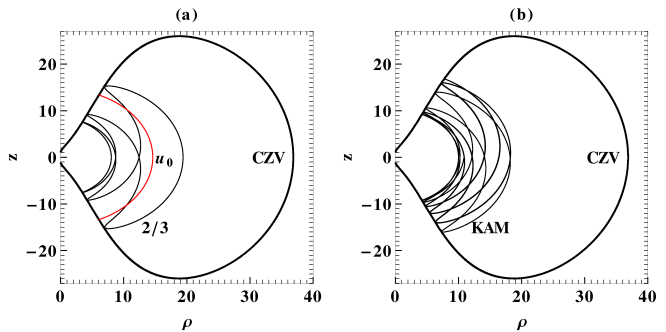}}
 \caption{The projections of geodesic orbits on the meridian plane ($(z,\rho)$)
 for $\delta=2,~k=1$, $E=0.95$ and $L_z=3$. (a) The simple periodic orbit
 $\mathbf{u}_0$ (red/gray curve) and the $2/3$ periodic orbit (black curve).
 (b) A quasiperiodic (KAM) orbit.
 In both panels, the boundary of motion on  the meridian plane is determined by
 the CZV (thick curve). 
 }
 \label{FigOrbd2}
\end{figure}

\begin{figure*}[htp]
 \centerline{\includegraphics[width=0.75\textwidth] {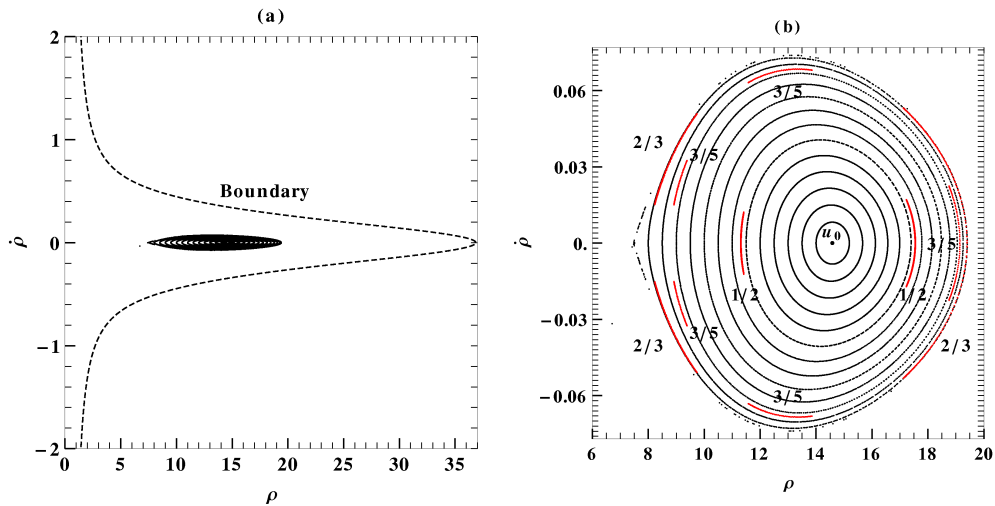}}
 \caption{(a) The surface of section $z=0$ ($\dot{z}>0$) for $\delta=2$,
 $E=0.95$ and $L_z=3$. The dashed black curve determines the boundary of the
 bounded orbits. The closed curves are KAM curves of the main island
 of stability, while the empty phase space between the the main island of
 stability and the boundary curve is occupied by plunging orbits. (b) A
 magnification of the main island of stability. The point $\mathbf{u}_0$ is the
 central fixed point. Various islands of stability forming Birkhoff chains are
 shown, labeled by their corresponding rotation numbers.       
 }
 \label{FigSSd2GV}
\end{figure*}

Previous numerical investigations of the ZV system \cite{Sota96,Brink08} have
led to a conjecture that this system is integrable. This led to attempts of
finding the missing integral of motion \cite{Brink08,KrugMat11}. In
\cite{KrugMat11}, however, the authors proved the non-existence of a broad
category of integrals of motion for the $\delta=2$ case. The parameter $\delta=2$
was used also in \cite{Brink08}, where the values $E=0.95$, $L_z=3$ and $k=1$
($\delta=M$) were chosen for the numerical examples. The same values are applied
in the subsequent section \ref{subsec:dgt1} in order to provide a straight
forward comparison with previous studies, while cases for different values of
$\delta$ are presented as well.

\subsection{Cases with $\delta>1$}\label{subsec:dgt1}

The periodic orbits on the meridian plane are bounded by the CZV (section
\ref{subsec:GMZV}). Two examples of periodic orbits projected on the meridian
plane are given in Fig. \ref{FigOrbd2}a. The first is the simple periodic
orbit $\mathbf{u}_0$ (red/gray curve in Fig. \ref{FigOrbd2}a), which bounces
along an arc between two points of the CZV (thick curves in Fig. \ref{FigOrbd2}).
The second is a stable periodic orbit of the resonance $2/3$ (black curve in Fig.
\ref{FigOrbd2}a), which bounces also between points of the CZV but follows 
a more complex path. The path is even more complicated in the case of a
quasiperiodic orbit (Fig. \ref{FigOrbd2}b), because the quasiperiodic orbits
cover densely their hosting KAM tori (section \ref{subsec:HamDyn}).

Fig. \ref{FigSSd2GV}, now, shows the surface of section $z=0$ ($\dot{z}>0$)
corresponding to orbits like those of Fig. \ref{FigOrbd2}. The orbits shown on
the surface of section plane ($\rho,~\dot{\rho}$) are in general non-equatorial.
The bounded orbits lie inside the boundary curve
$\dot{\rho}=\pm \sqrt{-2~V_\textrm{eff}(z=0)}$ (dashed curve in Fig.
\ref{FigSSd2GV}a). Most of them are plunging orbits and only a small main island
of stability of non-plunging orbits appears in the center of Fig.
\ref{FigSSd2GV}a.

\begin{figure}[htp]
 \centerline{\includegraphics[width=0.35\textwidth] {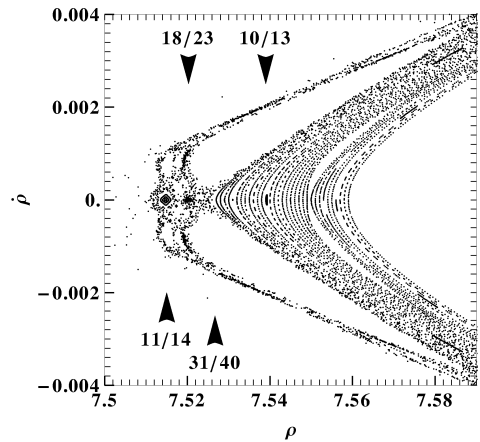}}
 \caption{Detail of the left tip of the main island of stability shown in Fig.
 \ref{FigSSd2GV}b. The arrows are showing the position of the islands of
 stability of the resonances $11/14,~18/23,~31/40,~10/13$ along the
 $\dot{\rho}=0$ line.
 }
 \label{FigSSd2d}
\end{figure}

At first glance the magnified surface of section in Fig. \ref{FigSSd2GV}b seems
to be quite regular, with no prominent signs of chaos. However, at the
resonances  $1/2,~3/5,~2/3$ thin islands of stability appear. These indicate the
existence of Birkhoff chains and, therefore, of chaos. In fact, if we zoom
at the left tip of the main island of stability we already get a typical picture
of a chaotic layer (Fig. \ref{FigSSd2d}). In Fig. \ref{FigSSd2d} the scattered
points define chaotic regions, while the continuous curves define the limits of
regular domains. However, the separation between chaotic and regular regions is
not so  clear. For example, the islands of stability belonging to the resonances
$11/14,~18/23,~31/40$ are embedded in prominent chaotic layers (Fig.
\ref{FigSSd2d}). 

Most of the chaotic orbits of Fig. \ref{FigSSd2d} are plunging orbits which,
due to the phenomenon of \emph{stickiness} \cite{ContopHars}, remain for a long
time close to regular non-plunging orbits, before crossing the LO and plunge.
The LO lies approximatively at the edge tip formed at the left of the main
island of stability of Fig. \ref{FigSSd2d}. In fact the region between the main
island of stability and the boundary curve (Fig. \ref{FigSSd2GV}a) is covered by
chaotic plunging orbits. The chaotic orbits stick around islands of higher
multiplicity, e.g. the islands of stability $11/14,~18/23,~31/40$ (Fig.
\ref{FigSSd2d}), before they plunge. The multiplicity is equal to the
denominator of the prime number ratio corresponding to the rotation number. 
        
\begin{figure}[htp]
 \centerline{\includegraphics[width=0.35\textwidth] {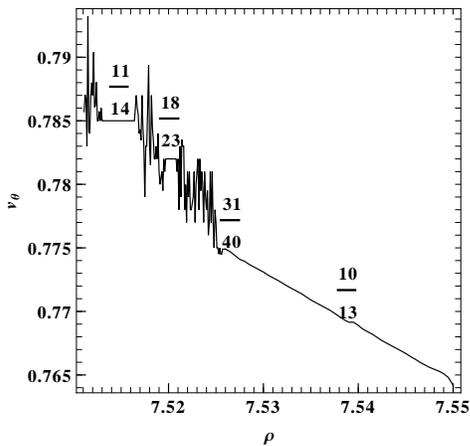}}
 \caption{The rotation number as a function of the initial conditions lying on
 the $\dot{\rho}=0$ line of Fig. \ref{FigSSd2d}. Prominent islands of stability
 are labeled by the respective rational value of $\nu_\theta$.
 }
 \label{FigZVd2Rotd}
\end{figure}

The rotation curve along the $\dot{\rho}=0$ line of Fig. \ref{FigSSd2d} is shown
in Fig. \ref{FigZVd2Rotd}. The regions dominated by regular motion are
represented by relatively smooth segments of the curve, while the chaotic
regions are recognized by various distinct parts where the rotation curve
develops fluctuations. At certain smooth segments of the rotation curve, the
rotation number is constant. These segments appear like ``plateaus'' of the
rotation curve. Distinct plateaus correspond to islands of stability of distinct
resonances in Fig. \ref{FigSSd2d}. In Fig. \ref{FigZVd2Rotd} the most prominent
plateaus are labeled by the value of the rotation number at each respective
resonance. 

\begin{figure}[htp]
 \centerline{\includegraphics[width=0.35\textwidth] {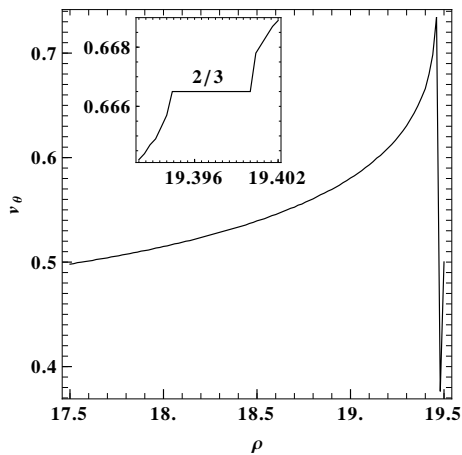}}
 \caption{The rotation curve along the line $\dot{\rho}=0$ corresponding to the
 Fig. \ref{FigSSd2GV}b. The embedded figure shows a detail of the curve at the
 $2/3$ resonance.
 }
 \label{FigZVd2RotGV}
\end{figure}

In fact, the computation of the rotation curve is an efficient way to look for
initial conditions leading to Birkhoff islands of stability, since the latter
can be located by the plateaus of the rotation curve. This method is
particularly useful in the location of tiny, or narrow islands, such us in
Fig. \ref{FigSSd2GV}b. In this case, by scanning along the line $\dot{\rho}=0$
the corresponding rotation curve (Fig. \ref{FigZVd2Rotd}) seems to be rather
smooth (excluding the last part corresponding to the chaotic layer surrounding
the main island of stability). However, a detailed scan reveals a number of
lower multiplicity resonances (e.g. the resonances $1/2,~2/3,~3/5$ shown in Fig.
\ref{FigSSd2GV}). Lower multiplicity resonant structures (like islands of
stability) are more prominent than higher multiplicity ones (as e.g. the
resonance $10/13$ shown in Fig. \ref{FigSSd2d}). By focusing on the lower
multiplicity resonances the anticipated plateaus of the islands of stability
clearly appear on the rotation curve. An example is shown for the plateau of
the conspicuous $2/3$ island of stability in the embedded panel of Fig.
\ref{FigZVd2RotGV}. 

\begin{figure}[htp]
 \centerline{\includegraphics[width=0.5\textwidth] {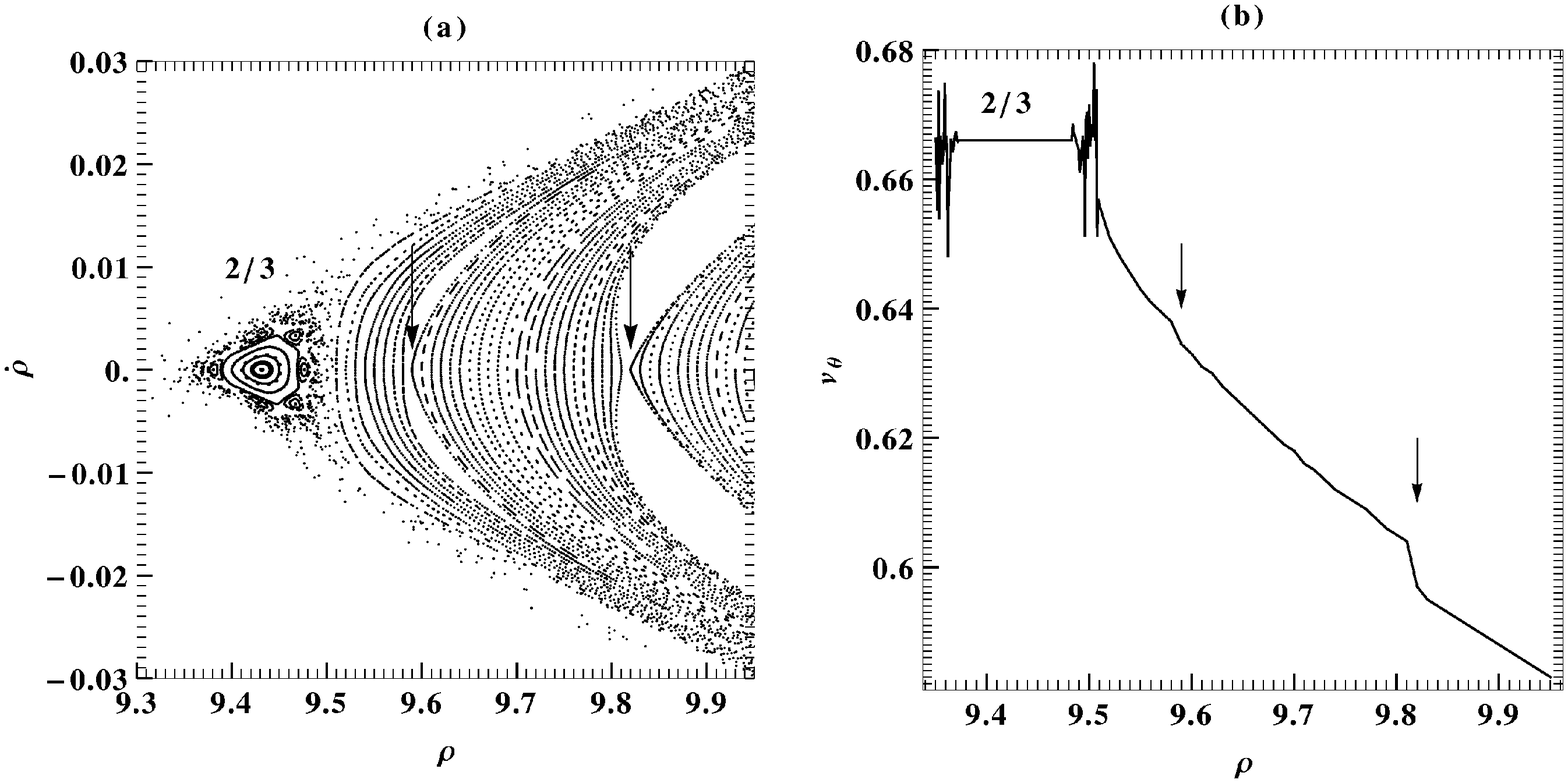}}
 \caption{(a) A detail of the $z=0$ ($\dot{z}>0$) surface of section for
 $\delta=3$, $E=0.98$, $L_z=7$ and $k=1$. An island of stability belonging to
 the $2/3$ resonance is shown embedded in a chaotic layer, while the arrows
 indicate the position of the unstable periodic points of other Birkhoff chains.
 (b) The rotation curve along the line $\dot{\rho}=0$ corresponding to Fig.
 \ref{FigZiVod3dV}a. The plateau of the $2/3$ is denoted and the arrows denote
 the behavior of the rotation curve when it crosses unstable periodic points. 
 }
 \label{FigZiVod3dV}
\end{figure}

If we change the value of the $\delta$ parameter to $\delta=3$ the structure of
the phase space does not change significantly. Near the outer boundary of the
main island there is a sticky chaotic layer of plunging orbits (Fig.
\ref{FigZiVod3dV}a) and Birkhoff chains appear inside the main island of
stability. The corresponding rotation curve in Fig. \ref{FigZiVod3dV}b shows
large variations whenever it crosses initial conditions belonging to chaotic
orbits. Also, the rotation curve takes the form of a plateau when crossing
resonant islands of stability, and it changes abruptly when crossing the
unstable periodic points of relatively small resonances (arrows in Fig.
\ref{FigZiVod3dV}b). Thus, in the case of $\delta=3$ as well both detecting
methods, i.e. the surface of section and the rotation number, indicate that the
ZV metric is not integrable.    

\begin{figure}[htp]
 \centerline{\includegraphics[width=0.5\textwidth] {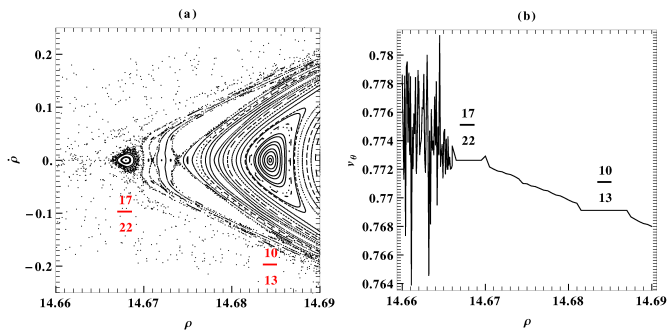}}
 \caption{(a) A detail of the $z=0$ ($\dot{z}>0$) surface of section for
 $\delta=4$, $E=0.95$, $L_z=3$ and $k=1$. (b) The rotation curve along the line
 $\dot{\rho}=0$ corresponding to Fig. \ref{FigZiVod4V}a.
 }
 \label{FigZiVod4V}
\end{figure}

If we further increase the value of $\delta$, the overall picture of the phase
space is qualitatively similar to the cases already presented. However, what
changes is the position of the unstable point ``x'' (section \ref{subsec:GMZV}),
which moves further away from the central anomaly $\rho=z=0$. This is expected,
since the position of the ISCO moves away as well (first equation of
eqs. (\ref{eq:Isco})). Thus, for increasing $\delta$ the main island of
stability moves all together to larger distances $\rho$. For example, Fig.
\ref{FigZiVod4V} shows a detail of the phase space for $\delta=4$, $E=0.95$,
$L_z=3$, where the whole island structure has been shifted further away from
$\rho=z=0$ with respect to Figs. \ref{FigSSd2d}, \ref{FigZVd2Rotd},
\ref{FigZiVod3dV}.

\begin{figure}[htp]
 \centerline{\includegraphics[width=0.5\textwidth] {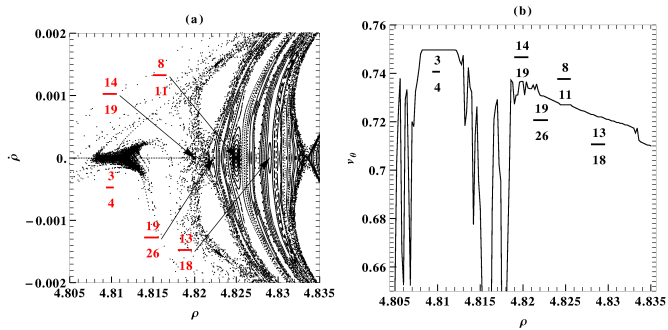}}
 \caption{(a) A detail of the $z=0$ ($\dot{z}>0$) surface of section for
 $\delta=3/2$, $E=0.98$, $L_z=5.7$ and $k=1$. (b) The rotation curve along the
 line $\dot{\rho}=0$ corresponding to Fig. \ref{FigZiVod3s2dV}a.
 }
 \label{FigZiVod3s2dV}
\end{figure}

In all previous cases the value of $\delta$ was a natural number. However, the
structure of the phase space seems not to change even if $\delta$ is a positive
real (not integer) number. For instance, if $\delta=3/2$, then we find again
a chaotic layer around the main island of stability (Fig. \ref{FigZiVod3s2dV}),
and all the complexity seen in the previous examples as well. In particular,
islands of stability are either embedded in a prominent chaotic layer or
enveloped between KAM curves (Fig. \ref{FigZiVod3s2dV}a). The resonances
corresponding to these islands are found in Fig. \ref{FigZiVod3s2dV}b by the use
of the rotation number. Fig. \ref{FigZiVod3s2dV}b is a good example to note
again that lower multiplicity islands of stability are more prominent than the
higher multiplicity ones. This fact is the reason why we expect
\cite{Apostolatos09,Lukes10} that the lower multiplicity islands of stability
are good candidates for detecting non-Kerr compact objects by the analysis of
gravitational waves coming from EMRIs, even if the inspiraling smaller compact
object might cross infinite resonances in a bumpy black hole spacetime
background during its inspiral.    

\subsection{Cases with $\delta<1$}\label{subsec:dlt1}

\begin{figure}[htp]
 \centerline{\includegraphics[width=0.5\textwidth] {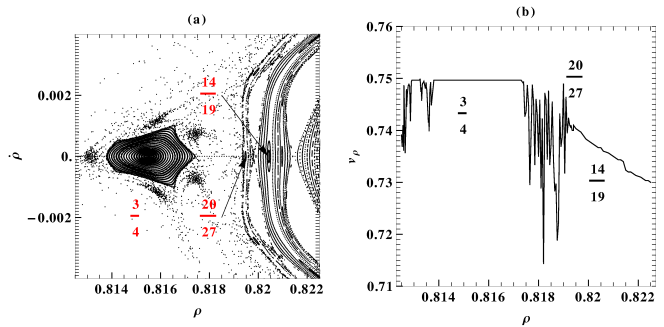}}
 \caption{(a) A detail of the $z=0$ ($\dot{z}>0$) surface of section for
 $\delta=1/2$, $E=0.99$, $L_z=1.8$ and $k=1$. (b) The rotation curve along the
 line $\dot{\rho}=0$ corresponding to Fig. \ref{FigZiVod1s2dV}a.
 }
 \label{FigZiVod1s2dV}
\end{figure}

When $\delta<1$, the ZV spacetime corresponds to a more prolate central object
than the respective Schwarzschild black hole. Nevertheless, for the prolate case
$\delta=1/2$ the structure of the phase space is similar to the oblate cases
examined in section \ref{subsec:dgt1}. There is a main island of stability, and
around it a chaotic sea of plunging orbits. In particular, the detail of the
surface of section $z=0$ ($\dot{z}>0$) (Fig. \ref{FigZiVod1s2dV}a) presents
a similar phase space structure near the sticky chaotic layer around the main
island of stability as in the oblate cases shown in section \ref{subsec:dgt1}.
Thus, we find again the $3/4$ Birkhoff chain, from which we can see clearly one
of the respective islands of stability embedded in the chaotic sea. Islands of
higher multiplicity lie near the border of the chaotic layer and of the main
island of stability. Thus, it seems that the ZV system has a similar dynamical
behavior independently of whether the corresponding central object is more
oblate or prolate than a Schwarzschild black hole.
 
\begin{figure}[htp]
 \centerline{\includegraphics[width=0.35\textwidth] {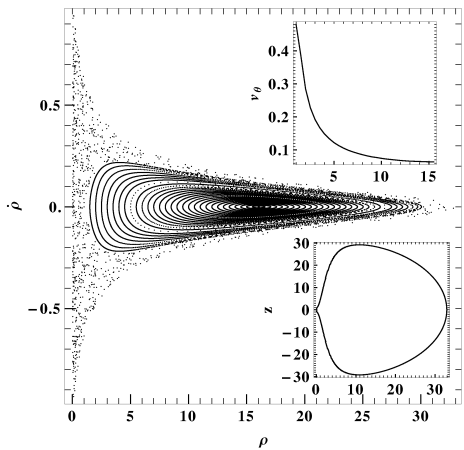}}
 \caption{The $z=0$ ($\dot{z}>0$) surface of section for $\delta=1/3$, $E=0.99$,
 $L_z=0.4$ and $k=1$. In the lower right embedded figure is shown the
 corresponding CZV. In the upper right embedded figure is shown the rotation curve
 along the $\dot{\rho}=0$ for the segment corresponding to the main island of
 stability. 
 }
 \label{FigZVd1s3TV}
\end{figure} 

However, if we reduce further the parameter $\delta$, for instance to
$\delta=1/3$, then the dynamical behavior changes significantly. This change is
expected for $\delta<1/\sqrt{5}$, because as noted in the last paragraph of
section \ref{subsec:GMZV} the ZV spacetime loses a property, namely the
existence of an ISCO, which is a characteristic of a black hole spacetime. In
Fig. \ref{FigZVd1s3TV} the surface of section shows a main island of stability
surrounded by a chaotic sea. Due to the fact that the CZV is a closed curve
(bottom right embedded panel in Fig. \ref{FigZVd1s3TV}), the orbits of the
chaotic sea cannot plunge towards the central anomaly. As a result, even if the
chaotic sea is prominent in size, no strong chaotic regions can be seen inside
the main island of stability. This fact is confirmed both by the surface of
section and by the rotation curve (main panel and right up panel in Fig.
\ref{FigZVd1s3TV} respectively). Thus, the structure of the main island of
stability seems not to change significantly for $\delta<1/\sqrt{5}$ (compare the
surface of section and the rotation curve in Fig. \ref{FigZVd1s3TV} with that in
Figs. \ref{FigSSd2GV}, \ref{FigZVd2RotGV} respectively).

\section{Conclusions and Discussion} \label{sec:ConDis}

It has been shown by clear numerical examples that the ZV metric corresponds in
general to a non-integrable system, contrary to the results of previous works
\cite{Sota96,Brink08}. The only known ZV spacetimes that correspond to
integrable systems, are the Minkowski spacetime ($\delta=0$) and the
Schwarzschild spacetime ($\delta=1$). In order to find imprints of the ZV
non-integrability the methods of the surface of section and of the rotation
number were employed.

In fact, in cases like the ZV system, we can only guess where to zoom on a
surface of section to discover chaotic layers or other imprints of chaos.
However, in such cases of nearly integrable Hamiltonian systems, chaos is
strongly correlated with the remnants of the destroyed resonance tori (section
\ref{subsec:HamDyn}). Therefore, a tool, like the rotation number, which detects
resonances is very useful.

Moreover, the rotation number is a reliable indicator of chaos. Thus, even if we
do not use a surface of section or another method for detecting chaos, the
rotation curves themselves are sufficient to show that chaos exists. This is
interesting also from an observational point of view, because the rotation
number is the ratio of the two main frequencies of a non-plunging geodesic orbit.
Therefore, the rotation number is an adequate tool for detecting phenomena
associated with chaos in gravitational wave signals. This has been already
studied in \cite{Apostolatos09,Lukes10}, where the rotation number was used in
order to detect deviations from the Kerr metric in a case of an EMRI into a
bumpy black hole spacetime background.

It has been shown that the ZV spacetime has a similar dynamical behavior in all
cases, i.e. independently of whether the spacetime corresponds to a more oblate
or prolate central compact object, as long as $\delta\ge 1/\sqrt{5}$. In
particular, all the orbits belonging to the chaotic sea which surrounds the main
island of stability are plunging. When, $0<\delta<1/\sqrt{5}$ then the ISCO
disappears and the orbits of the chaotic sea cease to be plunging. Thus,
$\delta=1/\sqrt{5}$ is a marginal value, where the ZV metric family changes its
behavior.

Finally, the study of bumpy black holes ISCOs has an interesting aspect
regarding spin measurements of Kerr black holes. If we suppose that ZV is
describing the spacetime around a compact object, then in the interval
$1/\sqrt{5}<\delta<1$ we get ISCO positions that correspond to the Kerr black
hole ones. This fact implies that if we evaluate the central compact object spin
by calculating the ISCO position like in \cite{McClintock11} without having
tested the Kerr hypothesis first, then these measurements might be misleading.

\begin{acknowledgments}
 I would like to thank V.~Matveev for steering my interest in this subject, and
 B.~Bruegmann, C.~Efthymiopoulos, C.~Markakis for usefull discussions and
 suggestions. G.~Lukes-Gerakopoulos was supported by the DFG grant
 SFB/Transregio 7.
\end{acknowledgments}

\appendix

\section{Numerical accuracy} \label{sec:NumAc}

\begin{figure}[htp]
 \centerline{\includegraphics[width=0.5\textwidth] {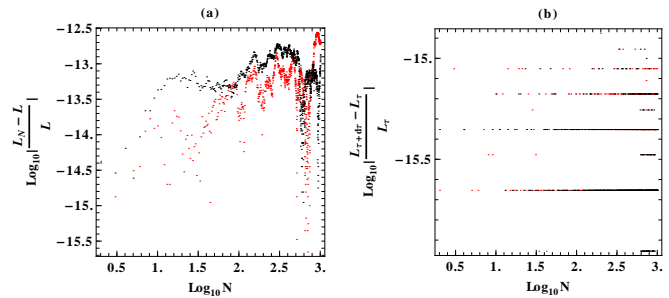}}
 \caption{(a) The relative error $|\frac{L_N-L}{L}|$ of the Lagrangian function
 $L$, which is a constant, as evaluated at the $N$th surface of section.
 (b) The relative error $|\frac{L_{\tau+d\tau}-L_\tau}{L_\tau}|$ between the
 Lagrangian function two consecutive integration steps at the $N$th surface of
 section. The red (gray) points corresponds to a chaotic orbit with initial
 condition $\dot{\rho}=z=0,~\rho=7.518$, while the black to a regular orbit with
 initial condition $\dot{\rho}=z=0,~\rho=7.548$. Both initial conditions lie on
 the surface of section shown in Fig. \ref{FigSSd2d}.
 }
 \label{FigReEr}
\end{figure} 

For integrating the geodesic orbits in the ZV system the 5th order Cash-Karp
Runge-Kutta was applied with a step size controller developed by the author. The
controller's function is to keep the relative error
$|\frac{L_{\tau+d\tau}-L_\tau}{L_\tau}|$
of the evaluated Lagrangian $L_\tau$, $L_{\tau+d\tau}$ functions between two
consecutive integration steps $\tau$, $\tau+d\tau$ below a tolerance value.
Even if such a control can secure the accuracy between two consecutive
integration steps, it cannot guarantee the accuracy of the overall calculation.
The integration accuracy is checked by the relative error $|\frac{L_\tau-L}{L}|$,
where the evaluated Lagrangian function $L_\tau$ is compared with the
theoretical one $L$.

The overall time evolution of the relative error $|\frac{L_N-L}{L}|$ at the Nth
surface of section crossing is shown in Fig. \ref{FigReEr}a for a regular orbit
(black dots) and for a chaotic orbit (red/gray points). The evolution of the
relative error $|\frac{L_{\tau+d\tau}-L_\tau}{L_\tau}|$ for the same orbits is
shown in Fig. \ref{FigReEr}b. Although the relative error at consecutive steps
appears rather small, the overall error shows that both chaotic and regular
orbits exhibit slow numerical drift. However, in both cases this drift is very
small compared to the scale of studied phenomena.

\end{document}